\documentclass[aps,prl,twocolumn,superscriptaddress,groupedaddress]{revtex4}  
\usepackage{graphicx}  
\usepackage{dcolumn}   
\usepackage{bm}        
\usepackage{amssymb}   
\usepackage{amsmath}   
\usepackage{mathtools} 
\usepackage{color}
\usepackage{ulem}

\usepackage[colorlinks,linkcolor=blue,urlcolor=blue,citecolor=blue]{hyperref}

\begin{document}

\def\be{\begin{equation}}
\def\ee{\end{equation}}
\def\bea{\begin{eqnarray}}
\def\eea{\end{eqnarray}}
\def\l{\label}

\newcommand{\eref}[1]{Eq.~(\ref{#1})}%
\newcommand{\Eref}[1]{Equation~(\ref{#1})}%
\newcommand{\fref}[1]{Fig.~\ref{#1}} %
\newcommand{\Fref}[1]{Figure~\ref{#1}}%
\newcommand{\sref}[1]{Sec.~\ref{#1}}%
\newcommand{\Sref}[1]{Section~\ref{#1}}%
\newcommand{\aref}[1]{Appendix~\ref{#1}}%
\newcommand{\sgn}[1]{\mathrm{sgn}({#1})}%
\newcommand{\erfc}{\mathrm{erfc}}%
\newcommand{\erf}{\mathrm{erf}}%

\title{Work fluctuations and Jarzynski equality in stochastic resetting}
\author{Deepak Gupta}
\affiliation{Dipartimento di Fisica `G. Galilei', INFN, Universit\'a di Padova, Via Marzolo 8, 35131 Padova, Italy}
\author{Carlos A. Plata}
\affiliation{Dipartimento di Fisica `G. Galilei', INFN, Universit\'a di Padova, Via Marzolo 8, 35131 Padova, Italy}

\author{Arnab Pal}
\thanks{Corresponding author}
\email{arnabpal@mail.tau.ac.il}
\affiliation{School of Chemistry, Raymond and Beverly Sackler Faculty of Exact Sciences, Tel Aviv University, Tel Aviv 6997801, Israel}
\affiliation{Center for the Physics and Chemistry of Living Systems. Tel Aviv University, 6997801, Tel Aviv, Israel}
\affiliation{The Sackler Center for Computational Molecular and Materials Science, Tel Aviv University, 6997801, Tel Aviv, Israel}

\date{\today}

\begin{abstract}
\noindent
We consider the paradigm of an overdamped Brownian particle in a potential well, which is modulated through an external protocol, in the presence of stochastic resetting. Thus, in addition to the 
short range diffusive motion, the particle also experiences intermittent long jumps which reset the particle back at a preferred location. Due to the modulation of the trap, work is done on the system 
and we investigate the statistical properties of the work fluctuations. We find that the distribution function of the work typically, in asymptotic times, converges to a universal Gaussian form for 
any protocol as long as that is also renewed after each resetting event. When observed for a finite time, we show that the system does not generically obey the Jarzynski equality which connects the 
finite time work fluctuations to the difference in free energy, albeit a restricted set of protocols which we identify herein. In stark contrast, the Jarzynski equality is always fulfilled when the protocols continue to evolve without being reset.
We present a set of exactly solvable models, demonstrate the validation of our theory and carry out numerical simulations to illustrate these findings. Finally, we have pointed out possible realistic implementations for resetting in experiments using the so-called engineered swift equilibration.

\end{abstract}

\maketitle

\textbf{Introduction.---}
Stochastic thermodynamics is a cornerstone in non-equilibrium statistical physics \cite{Sekimoto-ST,Seifert-review-08,Seifert-review-12,Jarzynski-review,FT-books}. Microscopic systems satisfy stochastic laws of motion governed by force fields 
and thermal fluctuations
which arise due to the surrounding. The subject then teaches us that thermodynamic observables such as work, heat, entropy production etc. measured along the stochastic trajectories taken from ensembles of such dynamics will fluctuate too.
Understanding the distribution and the statistical properties of these fluctuations is of great interest since they hold a treasure trove of information about microscopic systems and how they respond to external perturbations. Indeed
there has been a myriad of studies to understand e.g., 
non-equilibrium dynamics of biopolymers \cite{RitortRNA,JEexptLiphardt}, colloidal particles \cite{Seifert-non-harmonic,Ciliberto-JE1,Ciliberto-JE2,Imparato,Ciliberto-JE3,Ciliberto-JE4,workpal}, efficiency of molecular 
bio-motors \cite{efficiency-bio-1,efficiency-bio-2} and microscopic engines \cite{efficiency-engine-1}, heat conduction \cite{Lepri,Abhishek}, electronic transport in quantum systems \cite{Pekola1}, 
trapped-ion systems \cite{JEexption} and many more \cite{Ciliberto-JE5}. 
Although we observe such diverse small systems with no apparent similarity, it is remarkable to find that there exist some universal relations which are shared in common.
One of the most celebrated ones is perhaps the Jarzynski equality (JE) that relates the non-equilibrium fluctuations of the work to the equilibrium 
free energy difference \cite{JE97-1,JE97-2,JE97-3}. Universalities of such kind have always been considered as
an important feature in physical sciences and in this paper we seek out for thermodynamic 
invariant principles in stochastic resetting systems \cite{Restart1}.

Dynamics with stochastic reset has drawn a lot of attention recently because of its rich non-equilibrium properties \cite{Restart1,Restart2,Kirone,Restart-KPZ,transport1,transport2,Pal-potential,Pal-time-dep,invariance,invariance2,Edgar-path-integral,SRRW,SEP,underdamped,new-1,new-2} and its broad applicability in first passage processes \cite{ReuveniEnzyme,ReuveniPRL,PalReuveniPRL,branching,Landau,HRS,Restart-Search1,Chechkin,Belan,VV,reviewSR}. 
Nevertheless, thermodynamical perspective of resetting systems has been largely overlooked so far. 
It was only recently when first and second laws of thermodynamics were interpreted by identifying the contributions to the total entropy production \cite{thermo1}, and furthermore it was shown to satisfy a universal integral fluctuation relation \cite{thermo2}. While these first studies focused exclusively on the entropy production, efforts are yet to be made to understand other response functions. Moreover, not much is known about the distribution of these observables. 
In particular, one important observable is the work function which is produced due to external perturbations to the system. Work statistics encodes important features of an out-of-equilibrium thermodynamic process but its computation is usually
quite daunting. Here, we set out to characterize work fluctuations in a stochastic system which is subjected to resetting. Our detailed analysis to this account then reveals emergence of robust universal pattern in work-fluctuations: 
firstly resetting renders work-fluctuations Gaussian independent of the nature of the external perturbation that produces it. Secondly, work fluctuations are found to obey the JE under certain conditions which we identify through this
comprehensive study.

\textbf{General theory}.--- For the sake of generality, we put forward our results in the paradigmatic framework of a one-dimensional overdamped Brownian particle in a potential $U(x,\lambda(t))$, which is modulated externally through the protocol $\lambda(t)$. Motion of such a particle is governed by the Langevin equation of the form
\begin{equation}
\label{eq:Langevin}
 \dot{x}(t)=- \gamma^{-1}\partial_x U(x,\lambda(t)) + \sqrt{2D} \eta(t),
\end{equation}
where $\gamma$ and $D$ are the friction and diffusion coefficients respectively that satisfy the fluctuation-dissipation relation, i.e., $D\gamma=k_B T$, with $k_B$ being the Boltzmann constant and $T$ is the temperature of the medium. We assume $\langle \eta (t) \rangle =0$ and $\langle \eta (t)\eta(t') \rangle =\delta(t-t')$. Moreover, let us consider that the position of the particle at $t=0$ is distributed according to the probability density function (PDF) $p_{\text{ini}}(x_0)$. 
At random times taken from an exponential distribution
$f(t)=re^{-rt}$, the particle in motion is stopped and teleported to the initial configuration.

The external modulation of the potential performs work on the system which can be defined as \cite{JE97-1}
\begin{equation}
\label{eq:Wdef}
W=\frac{1}{k_BT} \int_0^t \mathrm{d}t' \  \dfrac{\partial U [x(t'),\lambda(t')]}{\partial \lambda}\frac{d\lambda(t')}{dt'}~,
\end{equation}
measured in units of $k_BT$. In what follows, we will set $k_BT=D=1$ without any loss of generality.

Since the reset process is instantaneous, we will assume that no work was done during this course (a finite resetting-work will be discussed later). 
In order to quantify the work fluctuations, it is convenient to first define the moment generating function (MGF) namely
\bea
\label{eq:Hrdefmain}
H_r(k,t) &\equiv &\int_{-\infty}^{\infty} \mathrm{d}W\ e^{-k W}P_r(W,t)
\eea
where $P_r(W,t)$ is the PDF of the work at time $t$,
averaged over the initial distribution $p_{\text{ini}}(x_0)$ and the underlying dynamics with stochastic resetting. To delve deeper, we make use of the renewal structure in resetting dynamics to
construct a relation that connects the MGF for $r>0$ to that of $r=0$ for any initial and subsequent resetting positions
\begin{equation}
\tilde{H}_r(k,s)=\frac{\tilde{H}_0(k,s+r)}{1-r \tilde{H}_0(k,s+r)},
\label{laplace-formula}
\end{equation} 
where $\tilde{H}_r(k,s)=\int_0^\infty~\mathrm{d}t~e^{-st}~H_r(k,t)$ and the subscript $0$ indicates the observables with $r=0$. 
We have added a proof of \eref{laplace-formula} in \cite{SM}; but it is imperative to stress the following points here. Note that \eref{laplace-formula} holds for any initial condition and naturally adheres to a fixed initial condition 
which was derived in \cite{restart_conc17,Hugo-Satya}, but in the absence of any protocol $\lambda(t)$. In the presence of protocol, 
one needs to be meticulous since the structure of this equation relies on the fact that $\lambda(t)$ \textit{is also renewed} after each resetting. As we will see later, \eref{laplace-formula} \textit{does not hold} when the protocol 
is \textit{unaffected} under resetting \cite{SM}.

The MGF, given by \eref{laplace-formula}, can be inverted
to obtain the full work statistics at a given time. Nonetheless, we will show that it suffices to know the first and second moment to predict the universal behavior of the work fluctuations in the large time limit. 
To this end, we first note that the $n$-th moment of $W$ in Laplace space can be written as $\int_0^\infty \mathrm{d}t e^{-st} \langle W^{n}(t) \rangle_r=\frac{\partial^n}{\partial(-k)^n}\tilde{H}_r(k,s)\big|_{k \to 0}$,
which satisfies a recursive-renewal structure \cite{SM}

{\footnotesize
\bea
\tilde{W}^n_r(s)=\frac{s+r}{s} \left[ \tilde{W}^n_0(s+r)+r\sum_{l=1}^n \begin{pmatrix}n\\l\end{pmatrix} \tilde{W}^{n-l}_r(s)~ \tilde{W}^l_0(s+r) \right],
\label{moments}
\eea}
where we have defined $\tilde{W}^n_r(s) \equiv \int_0^\infty~dt~e^{-st}~\langle W^{n}(t) \rangle_r$. \eref{moments} gives a simple recipe to compute all the moments of $W$ recursively from the knowledge of the moments of the process without resetting.

\begin{figure}
    \includegraphics[width=0.506\textwidth,height=5.5cm]{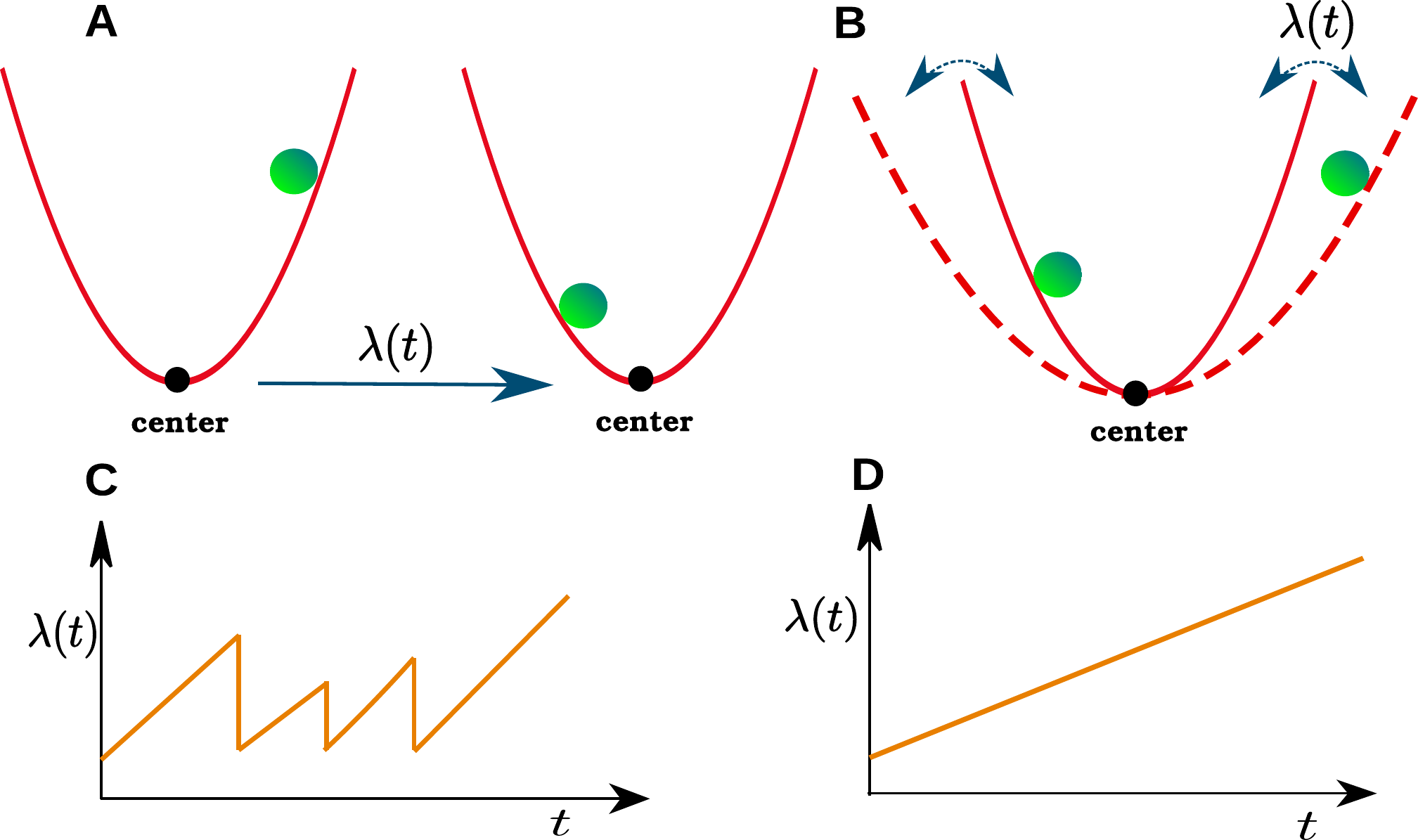}
    \caption{Schematic of a Brownian particle confined in a harmonic trap $U(x,\lambda(t))=\kappa(t)[x-y(t)]^2/2$, where $\lambda(t)=\{y(t),\kappa(t)\}$ represents the set of time-dependent protocols which are independently regulated. 
    $\lambda=y(t)$ and $\lambda=\kappa(t)$  indicate the center of the trap (panel A) and  the stiffness (panel B) respectively. The resetting mechanism acts both on the particle and the protocols as mentioned in the text. 
    Here, we show the modulation of the protocol when it is renewed after each resetting (panel C) or remains unaltered (panel D).}
   \label{Fig1}
\end{figure}

\begin{figure}[t]
  \begin{center}
     \includegraphics[width=8.75cm]{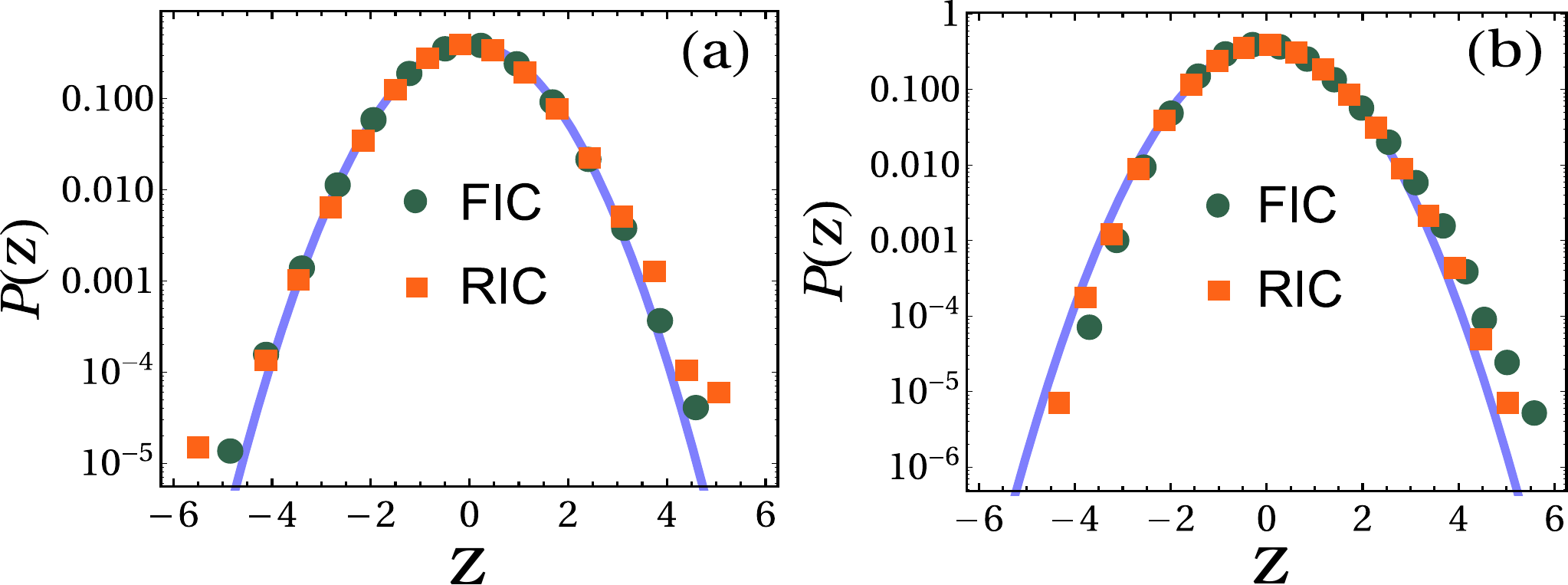}
    \caption{Numerical computation of the PDF of the rescaled work $z=(W-\mu_t)/\sigma_W(t)$ performed on a Brownian particle in a harmonic trap for the linear modulation of the trap center i.e., $y(t)=ut$ (panel a) and the stiffness i.e., $\kappa(t)=\kappa_0+vt$ (panel b) respectively. Simulations are performed for fixed initial condition (FIC): $p_{\text{ini}}(x_0)=\delta(x_0)$ (circle markers) and random initial condition (RIC): $p_{\text{ini}}(x_0)=p_{\text{eq}}(x_0)\propto \exp \left[ - U(x_0,\lambda(0))\right]$ (square markers) respectively for each of the above cases. Parameters for panel (a): $\kappa_0=1.5, u=0.2$ for FIC and $\kappa_0=0.5, u=0.5$ for RIC respectively where $r=0.5$ and $t=10$ are set identical for both of these cases. Similarly, parameters for panel (b): $\kappa_0=0.5, v=0.002, y=0, r=5$, and $t=500$ for both FIC and RIC. Numerical simulations are corroborated with the theoretical prediction (solid line in both cases) given by $P(z)=e^{-z^2/2}/\sqrt{2 \pi}$, and we see an excellent Gaussian collapse.} 
    \label{fig:pw}
  \end{center}
\end{figure}

\textbf{Universal work fluctuations.}---The infinite set of moments given by \eref{moments} contains the same information as that of the full distributions $P_r(W,t)$. 
However, physical intuition tells us that not all the moments contribute significantly at long time. To see this, we consider a trajectory of time length $t$ with multiple possible resetting events. 
The total work done along this long trajectory can then be decomposed into the sum of the partial works produced in each time interval between the resetting events. 
However, these intervals are statistically independent since the entire configuration of the system (comprising the particle and the trap) is renewed after each resetting event, and 
hence there are no correlations between the intervals. Therefore, for a long enough observation time $t$ one would expect on an average $\sim rt$ number of resetting events and the total work $W(t)$ can 
then be written as $W \approx W_1+W_2+W_3+\cdots+W_{[rt]} $. Since the intervals are disjoint, the $W_i$-s are also independent and identically distributed. Moreover, if $W_i$-s are regular (with finite mean and variance), 
one would expect that the distribution of $W$, according to the central limit theorem, would converge to a Gaussian irrespective of the nature of the potential and choice of the external protocol
\begin{align}
P_r(W,t)=\dfrac{1}{\sqrt{2 \pi \sigma_W^2(t)}} \exp\bigg[-\dfrac{(W-\mu_t )^2}{2\sigma_W^2(t)}\bigg]~,
\label{dist-w-r}
\end{align}
where the mean $\mu_t \equiv \langle  W \rangle_r$ and the variance $\sigma_W^2(t) \equiv \langle W^2 \rangle_r-\langle W \rangle_r^2$ are computed from \eref{moments}. In \fref{fig:pw},
we demonstrate \eref{dist-w-r} in the set up of a 1D Brownian particle confined in a harmonic trap $U(x,\lambda(t))=\kappa(t) [x-y(t)]^2/2$, where $\kappa(t)$ and $y(t)$ represent the stiffness and center of the trap respectively (leaving details of the simulation in \cite{SM}).

\textbf{Jarzynski equality---reset protocol}. The JE relates the finite time work fluctuations to the equilibrium free energy and here we ask whether such relations hold generically in resetting systems. We consider the same set-up as before and assume that each resetting act renews both the particle and the protocol. We further assume that the 
initial condition is taken from an equilibrium distribution  
$p_{\text{ini}}(x_0)\propto \exp \left[ - U(x_0,\lambda(0))\right]$, which is an essential
prerequisite for the JE. Employing \eref{laplace-formula} and substituting $k=1$ there, we find a renewal expression for the average of the exponentiated work which connects to the same with $r=0$ in the Laplace space \cite{SM}
\begin{align}
\mathcal{L}_{t\to s} \left[ \langle e^{-W}\rangle_r \right]
= \dfrac{\mathcal{L}_{t \to s+r}[\langle e^{-W}\rangle_0]}{1-r \mathcal{L}_{t \to s+r} [\langle e^{-W}\rangle_0]}~,
\label{je-r}
\end{align}
where $\mathcal{L}$ is the Laplace transform operator.
Several comments are in order now.
The exponential average on the RHS is along the trajectory without resetting and therefore must satisfy the JE i.e.,
$\langle e^{-W_{[0,t]}}\rangle_0=e^{- \left[ F_0(\lambda(t))-F_0(\lambda(0)) \right]}$, where $F_0(\lambda(t))$ is the free energy of the underlying system (i.e., when the dynamics is not interrupted by resetting) corresponding to the value of $\lambda$ evaluated at time $t$. However, it is evident that substituting this in \eref{je-r}
will not essentially lead to $ e^{- \Delta F_0(t)}$ (where $\Delta F_0(t)=F_0(\lambda(t))-F_0(\lambda(0))$ is the free-energy difference) along the entire trajectory of length $t$ in the presence of resetting i.e., JE will not be obeyed generically for any arbitrary protocol.  Nonetheless, we identify the protocols which will indeed satisfy this condition. This happens when the modulation of the protocol renders a linear change in the free energy i.e., $\Delta F_0(t)=\alpha t$. The trivial scenario i.e., $\Delta F_0(t)=0$ is true under any external perturbation which is of the following form: 
$U(x,y(t))=U(x-y(t))$. This could happen, e.g., when we move the center of the trap $y(t)$ according to some specific schedule. On the other hand, the nontrivial linear change in $\Delta F_0 (\neq 0)$ occurs when e.g., the stiffness $\kappa(t)$ is varied exponentially as a function of time. Utilizing this condition in \eref{je-r}, we obtain $\langle e^{-W} \rangle_r=e^{- \Delta F_0(t)}$ which holds along the entire trajectory with multiple resetting events \cite{SM}.

\begin{figure}[t]
  \begin{center}
  \includegraphics[width=8.75cm]{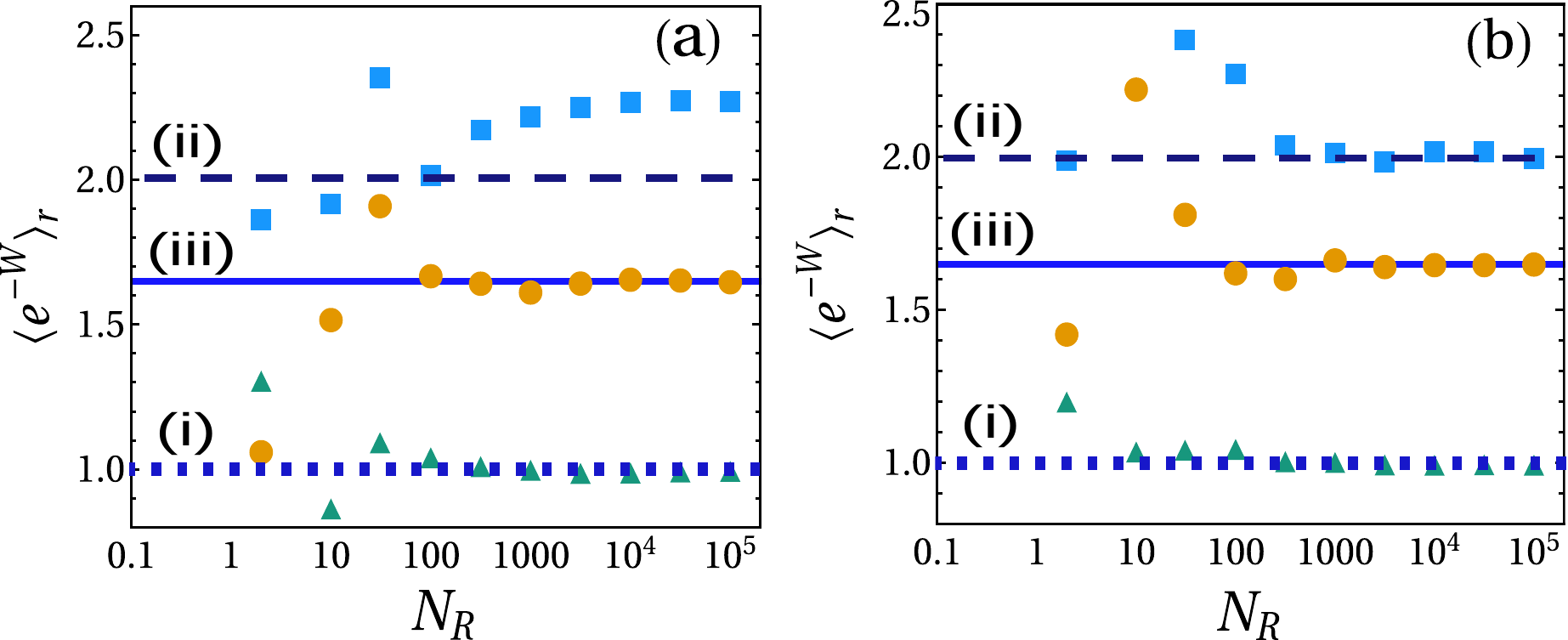}
    \caption{Numerical verification of the JE:
    we have demonstrated convergence of $\langle e^{-W} \rangle_r$ as a function of the number of realizations $N_R$. We have used three different types of protocol modulations as mentioned in the main text. 
    The analytical values of $e^{- \Delta F_0}$, shown by the horizontal lines (dotted for (i) moving trap, dashed and solid for the (ii) power law and (iii) exponential stiffness respectively in both panels), are plotted against
   numerical points for $\langle e^{-W} \rangle_r$ (shown by the triangles, squares, and circles respectively).
    Panel (a): Reset protocol. JE is seen to hold for protocols (i) and (iii) but not (ii). Parameters:  $\kappa_0=1.5$ and $\kappa_0=0.5$ respectively for the center and stiffness modulation.
  Panel (b): Non-reset protocol. JE holds for any protocols. Parameters: $\kappa_0=1$ and  $\kappa_0=0.35$ respectively for the center and stiffness modulation. In all the simulations, we have set $r=0.5$ and $t=5$. }
 \label{JE}
  \end{center}
\end{figure}

We now briefly summarize the numerical setups which are used to verify these findings.
We have simulated an overdamped Brownian particle in a harmonic trap $U(x,\lambda(t))=\kappa(t)[x-y(t)]^2/2$ in the presence of resetting ($r=0.5$), and measured $e^{-W}$ till time $t=5$. In \fref{JE}a, we have shown the convergence of the statistical average $\langle e^{-W} \rangle_r$ as a function of realizations $N_R$ for the following protocol modulations (i) moving the center of the trap with $y(t)=0.2t$, (ii) changing stiffness with a power law  $\kappa(t)=\kappa_0 (1+0.2t)^{-2}$, and (iii) an exponential law $\kappa(t)=\kappa_0 e^{- 0.2 t}$. As before,
we have regulated one protocol at a time keeping the others fixed. The horizontal lines shown in the panel correspond to the theoretical prediction of $e^{- \Delta F_0(t)}$ which takes the values $1.0$, $2.0$ and $\sim 1.65$ respectively for each of the modulations. The exact computation has been reserved to \cite{SM}. It is evident from \fref{JE}a that the JE holds for modulations (i) and (iii), but not for modulation (ii).

\textbf{Jarzynski equality is invariant under non-reset protocol}.---The discussion so far focused on the case when we reset both the protocol and the particle. In the following, we relax this condition and assume that only the particle is reset while the protocol keeps evolving in time.
Moreover, we consider that after each resetting event, position of the particle is drawn from the equilibrium distribution $p_{\text{ini}}(x_0)\propto \exp \left[ - U(x_0,\lambda(t_i))\right]$ corresponding to $\lambda$ measured at the times $t_i$ of resetting. In this way, the particle is \textit{effectively} equilibrated after each resetting event which is essential for the JE to hold. This construction correlates the intervals between resetting events: since the initial configuration of a given interval depends on the time spent in the previous one and hence renewal structure of \eref{laplace-formula} is lost \cite{SM}. 
However, notice that (i) the particle is prepared at the equilibrium state $p_{\text{ini}}$ after each resetting event, and (ii) consequently, the equality is satisfied in any interval between two resetting events. Taking these two facts into account, one can show that the equality holds along the entire trajectory independent of the nature of the protocol \cite{SM}
\bea
\label{eq:JE-extension}
\langle e^{-W} \rangle_r=e^{- \Delta F_0(t)}~.
\eea
We numerically check \eref{eq:JE-extension} in \fref{JE}b 
and show that indeed JE is invariant under \textit{non-reset protocol} modulations.

\textbf{Discussion.---} Up to this point, we have strictly assumed that resetting is an instantaneous process and thus neglected any possible contributions coming from it to the total work done. However, in real world taking one particle from location A to B will require work and this contribution must be taken into account.
Another essential aspect of the paper is the imposed \textit{equilibrated condition} upon each resetting which may appear artificial from an experimental point of view. To fill up these conceptual gaps, in this section, we put forward an explicit proposal for practical implementation of resetting which accounts for both these issues. To proceed further, recall that the unhindered process, characterized by $p_0(x,t)$, satisfies the Fokker-Planck equation
$
\partial_t p_0(x,t)= \partial_x \left[  \frac{1}{\gamma} U'(x,\lambda(t)) p_0(x,t)  \right] + D \partial_x^2 p_0(x,t).
$
Let us now focus on the first resetting event which occurs at a time $t_r$. Then the job of this engineered restart mechanism would be to take the current distribution $p_0(x,t_r)$ in the resetting time $t_r$ and make a transformation to reach an arbitrary target distribution $p_f(x)$ (which in our case is the equilibrium density) in \textit{a fixed time} $\Delta_r>0$. There have been recent developments to design protocols, namely Engineered Swift Equilibration (ESE) \cite{Trizac,Review-Guery,opt-seifert,opt-plata}, that shortcuts the relaxation times between two target distributions whose properties can be controlled in time. We will now show how to choose the optimal protocol that renders the average irreversible work during the resetting mechanism minimum.

As a representative case, we will consider the Brownian particle diffusing in a harmonic trap whose stiffness is timely modulated, that is, $U(x,t)=\kappa(t)x^2/2$. At time zero, the particle is prepared in equilibrium with a zero-mean Gaussian and standard deviation $\sigma(0)\equiv\sigma_0=\sqrt{\left\langle x^2(0) \right\rangle}$ with $\kappa(0)=\kappa_0$ which satisfies $\sigma_0^2=k_B T / \kappa_0$ from the equipartition theorem. Due to the nature of the potential, position density remains to be zero-mean Gaussian at all times i.e., $p_0(x,t)=\frac{e^{-x^2/2\sigma^2(t)}}{\sqrt{2\pi \sigma^2(t)}}$, where $\sigma(t)$ satisfies the following
evolution equation: $\sigma(t) \dot{\sigma}(t) = - \frac{\kappa(t)}{\gamma} \sigma^2(t)+D$ \cite{SM}.
Upon first restart at time $t_r$, the particle returns from a position which is distributed according to a  Gaussian distribution with $\sigma(t_r)$ (corresponding to $\kappa(t_r)$). The goal is then to \textit{design an optimal protocol} $\kappa(t)$  ($t_r<t<t_r+\Delta_r$) which drives the system from $\sigma(t_r)$ to $\sigma_0$ in time $\Delta_r$ within the most efficient energy consumption budget. The average work performed during this interval (with \textit{reset-protocol}) over many such trajectories is given by
\bea
\left\langle W_{rp} \right\rangle = \frac{1}{2} \int_{t_r}^{t_r+ \Delta_r} \mathrm{d}t~ \sigma^2(t) \dot{\kappa}(t) =W_1+\Delta F_r+ W^{\text{irr}}~,
\eea
where $W_1=\frac{1}{2} [\sigma_0^2 \kappa(0) - \sigma^2(t_r) \kappa(t_r) ]$, and $~\Delta F_r= - \frac{1}{2} k_BT \ln \left( \frac{\sigma_0^2}{\sigma^2(t_r)} \right)$ is the difference of free energy between the equilibrium states characterized by the initial and final variance \cite{SM}.  Finally, the third term $W^{\text{irr}}[\dot{\sigma}(t)]=\gamma \int_{t_r}^{t_r + \Delta_r} \mathrm{d}t~ \dot{\sigma}^2(t)$ can be identified as the irreversible work of the process which is always positive. Note that $W_1$ and $\Delta F_r$ are determined given the initial and target states leaving the dependence of the specific protocol and $\Delta_r$ only in $W^{\text{irr}}$. The optimal profile $\sigma_\text{opt}(t)$ that minimizes $W^{\text{irr}}[\dot{\sigma}(t)]$ can be immediately obtained using variational calculus and this reads \cite{SM}
\begin{equation}
\sigma_\text{opt}(t)= \sigma(t_r) + (t-t_r)\frac{ \sigma_0-\sigma(t_r)}{\Delta_r}~.
\end{equation}  
Substituting this into the evolution equation for $\sigma(t)$, one immediately finds the corresponding profile for the optimal protocol $\kappa_{\text{opt}}(t)$. Interestingly, $\kappa_{\text{opt}}(t)$ develops finite discontinuities at the threshold times i.e., $\kappa(t_r) \neq \underset{t \to t_r}{\lim} \kappa_{\text{opt}}(t) $ and $\kappa(t_r+\Delta_r)=\kappa(0) \neq \underset{t \to t_r +\Delta_r}{\lim} \kappa_{\text{opt}}(t)$, as was also found in other studies \cite{opt-seifert,opt-plata,Solon}. Implementing the protocol, we get the optimal irreversible work $W_{\text{opt}}^{\text{irr}}=\gamma \frac{(\sigma_0-\sigma(t_r))^2}{\Delta_r}$, which is exactly proportional to $\Delta_r^{-1}$ \cite{SM}. This ensues that a \textit{perfect} instantaneous resetting (i.e., the $\Delta_r \to 0$ limit) is not physically viable (if a work indeed is accounted for the entire resetting mechanism) since the energetic cost for each resetting jump will be divergent. A similar analysis follows also for the \textit{non-reset protocol} \cite{SM}.

\textbf{Conclusions and outlook.---}
In summary, this letter discusses statistical properties of work fluctuations in a stochastic resetting system. We find that the 
introduction of resetting renders the work fluctuations Gaussian in the large time for the \textit{reset protocols}. We infer that this is due to the renewal structure of the resetting process. Consequently, our approach also predicts emergence of Gaussian fluctuations for other thermodynamic observables such as dissipated heat, power flux or entropy production. A detailed analysis of this problem, however, remains to be seen. Furthermore, we note that only the \textit{typical fluctuations} become Gaussian as a fallout of the central limit theorem. On the other hand, it is only rational to believe that such universality of the fluctuations will be lost while looking at the tail behavior (atypical fluctuations) of the work-distribution. An outstanding challenge would be to extend our approach to capture such scenarios using large deviation theory \cite{HugoLDF}.

Our research also presents an extensive study on JE in resetting systems and
unravels different constraints on the temporal
behavior of the protocols to preserve the JE.
We have also put forward a neat feasible experimental pathway to implement real resets inspired by the recent developments on the edge between stochastic thermodynamics and control theory. We would like to stress the fact that the JE still prevails if the observable under study is the work $W$ (\eref{eq:Wdef}) performed only between the reset events, while the total work contribution namely $W+W_{rp}^{\text{tot}}$ (the latter averaged over all the resetting events) would lead to non-generic results specific to resetting mechanism.
Naturally, our study opens up a new research avenue in stochastic thermodynamics of resetting with
a great appeal to the experimental demonstration of the work fluctuation theorems in the resetting controlled biophysical \cite{Budnar} and single molecular systems using optical traps \cite{Trizac,single-molecule-1,single-molecule-2,single-molecule-4,single-molecule-5}.

\textbf{Acknowledgements.---} Deepak Gupta is supported by ``Excellence Project 2018'' of the Cariparo foundation. Carlos A.~Plata acknowledges the support from University of Padova  through  project STARS2018. Arnab Pal acknowledges support from the Raymond and Beverly Sackler Post-Doctoral Scholarship at Tel-Aviv University. Arnab Pal is indebted to Anupam Kundu, Shlomi Reuveni, and Urna Basu for illuminating discussions. 

\textbf{Author contributions.---} All the authors performed research and wrote the paper together.


\begin{thebibliography}{}

\bibitem{Sekimoto-ST} Sekimoto, K., 1998. Langevin equation and thermodynamics. Progress of Theoretical Physics Supplement, 130, pp.17-27.



\bibitem{Seifert-review-08} Seifert, U., 2008. Stochastic thermodynamics: principles and perspectives. The European Physical Journal B, 64(3-4), pp.423-431.

\bibitem{Seifert-review-12} Seifert, U., 2012. Stochastic thermodynamics, fluctuation theorems and molecular machines. Reports on progress in physics, 75(12), p.126001.


\bibitem{Jarzynski-review} Jarzynski, C., 2011. Equalities and inequalities: Irreversibility and the second law of thermodynamics at the nanoscale. Annu. Rev. Condens. Matter Phys., 2(1), pp.329-351.


\bibitem{FT-books} Klages, R., Just, W. and Jarzynski, C. eds., 2013. Nonequilibrium statistical physics of small systems. Wiley-VCH Verlag GmbH.

\bibitem{RitortRNA} Collin, D., Ritort, F., Jarzynski, C., Smith, S.B., Tinoco Jr, I. and Bustamante, C., 2005. Verification of the Crooks fluctuation theorem and recovery of RNA folding free energies. Nature, 437(7056), p.231.


\bibitem{JEexptLiphardt} Liphardt, J., Dumont, S., Smith, S.B., Tinoco, I. and Bustamante, C., 2002. Equilibrium information from nonequilibrium measurements in an experimental test of Jarzynski's equality. Science, 296(5574), pp.1832-1835.


\bibitem{Seifert-non-harmonic} Blickle, V., Speck, T., Helden, L., Seifert, U. and Bechinger, C., 2006. Thermodynamics of a colloidal particle in a time-dependent nonharmonic potential. Physical review letters, 96(7), p.070603.


\bibitem{Ciliberto-JE1}
Douarche, F., Ciliberto, S., Petrosyan, A. and Rabbiosi, I., 2005. An experimental test of the Jarzynski equality in a mechanical experiment. EPL (Europhysics Letters), 70(5), p.593.


\bibitem{Ciliberto-JE2}
Gomez-Solano, J.R., Petrosyan, A., Ciliberto, S., Chetrite, R. and Gawedzki, K., 2009. Experimental verification of a modified fluctuation-dissipation relation for a micron-sized particle in a nonequilibrium steady state. Physical review letters, 103(4), p.040601.

\bibitem{Imparato}
Imparato, A., Peliti, L., Pesce, G., Rusciano, G. and Sasso, A., 2007. Work and heat probability distribution of an optically driven Brownian particle: Theory and experiments. Physical Review E, 76(5), p.050101.

\bibitem{Ciliberto-JE3}
Gomez-Solano, J.R., Bellon, L., Petrosyan, A. and Ciliberto, S., 2010. Steady-state fluctuation relations for systems driven by an external random force. EPL (Europhysics Letters), 89(6), p.60003.

\bibitem{Ciliberto-JE4}
Jop, P., Petrosyan, A. and Ciliberto, S., 2008. Work and dissipation fluctuations near the stochastic resonance of a colloidal particle. EPL (Europhysics Letters), 81(5), p.50005.

\bibitem{workpal}
Pal, A. and Sabhapandit, S., 2013. Work fluctuations for a Brownian particle in a harmonic trap with fluctuating locations. Physical Review E, 87(2), p.022138.

\bibitem{efficiency-bio-1}
Camunas-Soler, J., Alemany, A. and Ritort, F., 2017. Experimental measurement of binding energy, selectivity, and allostery using fluctuation theorems. Science, 355(6323), pp.412-415.


\bibitem{efficiency-bio-2}
Alemany, A., Mossa, A., Junier, I. and Ritort, F., 2012. Experimental free-energy measurements of kinetic molecular states using fluctuation theorems. Nature Physics, 8(9), p.688.

\bibitem{efficiency-engine-1} Mart\'inez, I.A., Rold\'an, \'E., Dinis, L., Petrov, D., Parrondo, J.M. and Rica, R.A., 2016. Brownian carnot engine. Nature physics, 12(1), p.67.


\bibitem{Lepri} Lepri, S. ed., 2016. Thermal transport in low dimensions: from statistical physics to nanoscale heat transfer (Vol. 921). Springer.

\bibitem{Abhishek}
Saito, K. and Dhar, A., 2007. Fluctuation theorem in quantum heat conduction. Physical Review Letters, 99(18), p.180601.

\bibitem{Pekola1}
Saira, O.P., Yoon, Y., Tanttu, T., Mottonen, M., Averin, D.V. and Pekola, J.P., 2012. Test of the Jarzynski and Crooks fluctuation relations in an electronic system. Physical review letters, 109(18), p.180601.


\bibitem{JEexption}
An, S., Zhang, J.N., Um, M., Lv, D., Lu, Y., Zhang, J., Yin, Z.Q., Quan, H.T. and Kim, K., 2015. Experimental test of the quantum Jarzynski equality with a trapped-ion system. Nature Physics, 11(2), p.193.

\bibitem{Ciliberto-JE5}
Douarche, F., Joubaud, S., Garnier, N.B., Petrosyan, A. and Ciliberto, S., 2006. Work fluctuation theorems for harmonic oscillators. Physical review letters, 97(14), p.140603.

\bibitem{JE97-1} Jarzynski, C., 1997. Nonequilibrium equality for free energy differences. Physical Review Letters, 78(14), p.2690.

\bibitem{JE97-2} Jarzynski, C., 1997. Equilibrium free-energy differences from nonequilibrium measurements: A master-equation approach. Physical Review E, 56(5), p.5018.


\bibitem{JE97-3} Jarzynski, C., 2006. Rare events and the convergence of exponentially averaged work values. Physical Review E, 73(4), p.046105.



\bibitem{Restart1} Evans, M.R. and Majumdar, S.N., 2011. Diffusion with stochastic resetting. Physical review letters, 106(16), p.160601.

\bibitem{Restart2} Evans, M.R. and Majumdar, S.N., 2011. Diffusion with optimal resetting. Journal of Physics A: Mathematical and Theoretical, 44(43), p.435001.

\bibitem{Kirone}
Evans, M.R., Majumdar, S.N. and Mallick, K., 2013. Optimal diffusive search: nonequilibrium resetting versus equilibrium dynamics. Journal of Physics A: Mathematical and Theoretical, 46(18), p.185001.

\bibitem{Restart-KPZ}
Gupta, S., Majumdar, S.N. and Schehr, G., 2014. Fluctuating interfaces subject to stochastic resetting. Physical review letters, 112(22), p.220601.

\bibitem{transport1}
Majumdar, S.N., Sabhapandit, S. and Schehr, G., 2015. Dynamical transition in the temporal relaxation of stochastic processes under resetting. Physical Review E, 91(5), p.052131.

\bibitem{transport2}
Eule, S. and Metzger, J.J., 2016. Non-equilibrium steady states of stochastic processes with intermittent resetting. New Journal of Physics, 18(3), p.033006.

\bibitem{Pal-potential} Pal, A., 2015. Diffusion in a potential landscape with stochastic resetting. Physical Review E, 91(1), p.012113.


\bibitem{Pal-time-dep}Pal, A., Kundu, A. and Evans, M.R., 2016.
Diffusion under time-dependent resetting. Journal of Physics A: Mathematical
and Theoretical, 49(22), p.225001.


\bibitem{Edgar-path-integral} Rold\'an, \'E. and Gupta, S., 2017. Path-integral formalism for stochastic resetting: Exactly solved examples and shortcuts to confinement. Physical Review E, 96(2), p.022130.

\bibitem{invariance}
Pal, A., Ku\'smierz, \L{} and Reuveni, S., 2019. Invariants of motion with stochastic resetting and space-time coupled returns. New Journal of Physics, 21(11), p.113024.



\bibitem{invariance2}
Pal, A., Ku\'smierz, \L{} and Reuveni, S., 2019. Time-dependent density of diffusion with stochastic resetting is invariant to return speed. Physical Review E, 100(4), p.040101.


\bibitem{SRRW} M\'endez, V. and Campos, D., 2016. Characterization of stationary states in random walks with stochastic resetting. Physical Review E, 93(2), p.022106.

\bibitem{SEP}Basu, U., Kundu, A. and Pal, A., 2019. Symmetric exclusion process under stochastic resetting. Physical Review E, 100(3), p.032136.


\bibitem{underdamped}
Gupta, D., 2019. Stochastic resetting in underdamped Brownian motion. Journal of Statistical Mechanics: Theory and Experiment, 2019(3), p.033212.


\bibitem{new-1} Evans, M.R., Majumdar, S.N. and Mallick, K., 2013. Optimal diffusive search: nonequilibrium resetting versus equilibrium dynamics. Journal of Physics A: Mathematical and Theoretical, 46(18), p.185001.


\bibitem{new-2} Evans, M.R., Majumdar, S.N. and Schehr, G., 2019. Stochastic Resetting and Applications. arXiv preprint arXiv:1910.07993.



\bibitem{ReuveniEnzyme}
Reuveni, S., Urbakh, M. and Klafter, J., 2014. Role of substrate unbinding in Michaelis–Menten enzymatic reactions. Proceedings of the National Academy of Sciences, 111(12), pp.4391-4396.

\bibitem{ReuveniPRL}Reuveni, S., 2016. Optimal stochastic restart renders fluctuations in first passage times universal. Physical review letters, 116(17), p.170601.

\bibitem{PalReuveniPRL} Pal, A. and Reuveni, S., 2017. First Passage under Restart. Physical review letters, 118(3), p.030603.

\bibitem{branching}
Pal, A., Eliazar, I. and Reuveni, S., 2019. First passage under restart with branching. Physical review letters, 122(2), p.020602.

\bibitem{Landau}
Pal, A. and Prasad, V.V., 2019. Landau-like expansion for phase transitions in stochastic resetting. Physical Review Research, 1(3), p.032001.


\bibitem{HRS}
Pal, A., Ku\'smierz, \L{} and Reuveni, S., 2019. Home-range search provides advantage under high uncertainty. arXiv preprint arXiv:1906.06987.


\bibitem{Restart-Search1} Kusmierz, L., Majumdar, S.N., Sabhapandit, S. and Schehr, G., 2014. First order transition for the optimal search time of L\'evy flights with resetting. Physical review letters, 113(22), p.220602.

\bibitem{Chechkin}Chechkin, A. and Sokolov, I.M., 2018. Random search with resetting: a unified renewal approach. Physical review letters, 121(5), p.050601.


\bibitem{Belan} Belan, S., 2018. Restart could optimize the probability of success in a Bernoulli trial. Physical review letters, 120(8), p.080601.


\bibitem{VV} Pal, A. and Prasad, V.V., 2019. First passage under stochastic resetting in an interval. Physical Review E, 99(3), p.032123.

\bibitem{reviewSR} Evans, M.R., Majumdar, S.N. and Schehr, G., 2019. Stochastic Resetting and Applications. arXiv preprint arXiv:1910.07993.

\bibitem{thermo1} Fuchs, J., Goldt, S. and Seifert, U., 2016. Stochastic thermodynamics of resetting. EPL (Europhysics Letters), 113(6), p.60009.

\bibitem{thermo2}
Pal, A. and Rahav, S., 2017. Integral fluctuation theorems for stochastic resetting systems. Physical Review E, 96(6), p.062135.


\bibitem{SM} Gupta, D., Plata, C. A., and Pal, A. See Supplemental material.

\bibitem{restart_conc17} Meylahn, J.M., Sabhapandit, S. and Touchette, H., 2015. Large deviations for Markov processes with resetting. Physical Review E, 92(6), p.062148.

\bibitem{Hugo-Satya} Den Hollander, F., Majumdar, S.N., Meylahn, J.M. and Touchette, H., 2019. Properties of additive functionals of Brownian motion with resetting. Journal of Physics A: Mathematical and Theoretical, 52(17), p.175001.


\bibitem{Trizac} Martínez, I.A., Petrosyan, A.,  Gu\'ery-Odelin, D., Trizac, E. and Ciliberto, S., 2016. Engineered swift equilibration of a Brownian particle. Nature physics, 12(9), p.843.

\bibitem{Review-Guery} Gu\'ery-Odelin, D., Ruschhaupt, A., Kiely, A., Torrontegui, E., Mart\'inez-Garaot, S. and Muga, J. G., 2019. Shortcuts to adiabaticity: Concepts, methods, and applications. Review of Modern Physics, 91,  045001.

\bibitem{opt-seifert}Schmiedl, T. and Seifert, U., 2007. Optimal finite-time processes in stochastic thermodynamics. Physical review letters, 98(10), p.108301.  

\bibitem{opt-plata}Plata, C. A., Gu\'ery-Odelin, D., Trizac, E., and Prados, A., 2019. Optimal work in a harmonic trap with bounded stiffness. Physical Review E, 99, 012140.


\bibitem{Solon}
Solon, A.P. and Horowitz, J.M., 2018. Phase transition in protocols minimizing work fluctuations. Physical review letters, 120(18), p.180605.

\bibitem{HugoLDF}
Touchette, H., 2009. The large deviation approach to statistical mechanics. Physics Reports, 478(1-3), pp.1-69.

\bibitem{Budnar}Budnar, S., Husain, K.B., Gomez, G.A., Naghibosadat, M., Varma, A., Verma, S., Hamilton, N.A., Morris, R.G. and Yap, A.S., 2019. Anillin promotes cell contractility by cyclic resetting of RhoA residence kinetics. Developmental cell, 49(6), pp.894-906.

\bibitem{single-molecule-1}
Berut, A., Imparato, A., Petrosyan, A. and Ciliberto, S., 2016. Stationary and transient fluctuation theorems for effective heat fluxes between hydrodynamically coupled particles in optical traps. Physical review letters, 116(6), p.068301.


\bibitem{single-molecule-2}
Hoang, T.M., Pan, R., Ahn, J., Bang, J., Quan, H.T. and Li, T., 2018. Experimental test of the differential fluctuation theorem and a generalized Jarzynski equality for arbitrary initial states. Physical review letters, 120(8), p.080602.


\bibitem{single-molecule-4}
Admon, T., Rahav, S. and Roichman, Y., 2018. Experimental realization of an information machine with tunable temporal correlations. Physical review letters, 121(18), p.180601.

\bibitem{single-molecule-5}
Khan, M. and Sood, A.K., 2011. Irreversibility-to-reversibility crossover in transient response of an optically trapped particle. EPL (Europhysics Letters), 94(6), p.60003.


\end{thebibliography}
\end{document}